%% file: ssmpe.tex
\documentclass{article}
\usepackage[T1]{fontenc}
\usepackage[utf8]{inputenc}
\usepackage[]{ismir}
\usepackage{amsmath,cite,url}
\usepackage{graphicx}
\usepackage{xcolor}
\usepackage{color}
\usepackage{multicol}
\usepackage{threeparttable}

\title{Investigating an Overfitting and Degeneration Phenomenon in Self-Supervised Multi-Pitch Estimation}

\oneauthor
  {Frank Cwitkowitz \quad Zhiyao Duan}
  {Audio Information Research Lab, University of Rochester \\
   {\normalsize \tt fcwitkow@ur.rochester.edu, zhiyao.duan@rochester.edu}}

\def\authorname{F. Cwitkowitz and Z. Duan}

\begin{document}

\maketitle

\begin{abstract}
Multi-Pitch Estimation (MPE) continues to be a sought after capability of Music Information Retrieval (MIR) systems, and is critical for many applications and downstream tasks involving pitch, including music transcription. However, existing methods are largely based on supervised learning, and there are significant challenges in collecting annotated data for the task. Recently, self-supervised techniques exploiting intrinsic properties of pitch and harmonic signals have shown promise for both monophonic and polyphonic pitch estimation, but these still remain inferior to supervised methods. In this work, we extend the classic supervised MPE paradigm by incorporating several self-supervised objectives based on pitch-invariant and pitch-equivariant properties. This joint training results in a substantial improvement under closed training conditions, which naturally suggests that applying the same objectives to a broader collection of data will yield further improvements. However, in doing so we uncover a phenomenon whereby our model simultaneously overfits to the supervised data while degenerating on data used for self-supervision only. We demonstrate and investigate this and offer our insights on the underlying problem.
\end{abstract}

\section{Introduction}\label{sec:introduction}
Pitch is a perceptual attribute of sound events that produce waves or harmonics that oscillate at integer multiples of a fundamental frequency (F0) \cite{muller2015fundamentals}.
Pitch is a foundational aspect of music, and it is often useful to represent musical content in terms of relationships between pitch (\textit{i.e.}, melody and harmony).
In Music Information Retrieval (MIR) research, the task of detecting pitch activity and estimating the corresponding F0s within a polyphonic signal is known as Multi-Pitch Estimation (MPE) \cite{benetos2018automatic}.
This is an important task with exciting applications in machine listening, human-computer interaction, and music databasing.
Pitch estimation is also necessary for more high-level MIR tasks such as Automatic Music Transcription (AMT), where MPE is often performed in conjunction with the estimation of note events.
Currently, state-of-the-art MPE methods are heavily based on supervised machine learning techniques and require large amounts of rich and diverse training data with pitch annotations \cite{gardner2021mt3}.
However, there are significant challenges with procuring multi-pitch annotations, especially for audio recordings comprising multi-instrument mixtures, difficult-to-annotate polyphonic instruments (\textit{e.g.}, guitar), or less common instruments.
For this reason, such methods are unable to scale beyond the available datasets, which are generally homogeneous (\textit{e.g.}, solo piano) or limited in size (\textit{i.e.}, less than 10 hours).

Several strategies have been proposed to mitigate such issues, including semi-supervision on data with weakly aligned annotations \cite{maman2022unaligned} or pre-training on mixtures of large-scale monophonic data \cite{simon2022scaling}.
However, these methods still fall within the supervised learning paradigm and as such are subject to the size and quality of the data and annotations.
In parallel, there has also been work to build music foundation models that learn more general representations of music which can be transferred to downstream tasks \cite{li2024mert, liao2024music}.
However, these representations still struggle to capture the level of granularity needed for low-level tasks like MPE.
An alternative approach is to define task-specific self-supervised objectives that encourage a model to respect properties of pitch, such as equivariance to pitch shifting and invariance to timbral transformations \cite{gfeller2020spice, riou2023pesto}.
These techniques have demonstrated remarkable success in learning to estimate pitch from unlabeled data and have also been generalized to polyphonic data \cite{cwitkowitz2024toward}.

In this work, we expand upon, refine, and integrate the techniques proposed in \cite{cwitkowitz2024toward} into a supervised MPE framework resembling that of recent methods employing a convolutional neural network (CNN) \cite{bittner2017deep, bittner2022lightweight, weiss2022comparing, cwitkowitz2024timbre} trained to estimate a multi-pitch salience-gram for an input spectrogram.
We show these self-supervised objectives can significantly improve the performance of the supervised framework under a joint training paradigm.
However, when attempting to apply the same objectives to additional data with no corresponding supervision, we observe a surprising phenomenon: self-supervision on the additional data does not improve performance but actually steers our model toward degeneration, \textit{i.e.} blank pitch salience estimates, on such data.
The model still exhibits the correct behavior on the validation set of the annotated dataset as well as evaluation data following a similar distribution.
We demonstrate this issue and conduct several follow-up experiments in an attempt to identify and explain the underlying problem.

\section{Framework}\label{sec:framework}
In this section, we describe our feature extraction module, model architecture, and training objectives.
Our methodology can be viewed as the integration of self-supervised techniques for MPE \cite{cwitkowitz2024toward} into a supervised framework.

\subsection{Model \& Features}\label{sec:model_features}
We adopt a modified version of the fully convolutional 2D autoencoder used in the Timbre-Trap framework \cite{cwitkowitz2024timbre}.
This model comprises four encoder and decoder blocks with dilated convolutions, residual connections, and strided or transposed convolutions for resampling features across frequency.
Although Timbre-Trap was proposed as a unified framework to perform transcription and reconstruction, we adopt only the base model and discard the latent feature used to switch between modes.
We also insert layer normalization after the initial convolutional layer of both the encoder and decoder, after the strided and transposed convolution in each encoder and decoder block, and after the latent space convolution.
Finally, we double the number of filters in each convolutional layer.

We also replace the complex Constant-Q Transform (CQT) module used in the Timbre-Trap framework with calculation of Harmonic CQT\footnote{As in \cite{cwitkowitz2024toward}, a variable Q-factor \cite{schorkhuber2014matlab} is employed for improved computational efficiency and increased time resolution at lower frequencies.} (HCQT) spectrograms \cite{bittner2017deep} $X_\mathcal{H} \in [0, 1]^{6 \times K \times N}$ with $K = 440$ frequency bins starting from $f_{min} = 27.5$ Hz and $5$ bin per semitone resolution.
Input audio is resampled to $22,050$ Hz, and $N$ is the number frames using a hop size of $256$ samples.
We maintain the original set of harmonics $\mathcal{H} = \{0.5, 1, 2, 3, 4, 5\}$.
The main advantage of the HCQT is its capacity to index harmonic energy across the channel dimension.
This structure is perfectly suited for convolutional layers and establishes a strong inductive bias for pitch estimation.
One consequence of our model, denoted by $\mathcal{F}(\cdot)$, is the resulting shared dimensionality between $X_{\mathcal{H}}$ and predictions $\hat{Y} = \mathcal{F}\hspace{-0.15em}\left( X_{\mathcal{H}} \right)$, which ideally represent multi-pitch salience-grams.
This makes it convenient to formulate the self-supervised techniques proposed in Sec. \ref{sec:self_supervised_techniques}.
Note that this configuration of model and features is very similar to what was used in the SS-MPE framework \cite{cwitkowitz2024toward}.

\subsection{Supervised Training}\label{sec:supervised_training}
Given the ground-truth pitch activations $Y \in [0, 1]^{K \times N}$ corresponding to $X_{\mathcal{H}}$, a supervised loss can be defined as
\begin{equation}
\label{eq:supervised_loss}
\mathcal{L}_{spv} = \frac{1}{N} \sum^{N - 1}_{n = 0} \sum^{K - 1}_{k = 0} \mathcal{B}\hspace{-0.15em}\left( \hat{Y}[k, n], \widetilde{Y}[k, n] \right),
\end{equation}
where $\mathcal{B}(\cdot, \cdot)$ represents binary cross-entropy (BCE) loss and $\widetilde{Y}$ represents the target multi-pitch salience-gram for $X_\mathcal{H}$.
Following \cite{bittner2017deep}, we blur each frame of $Y$ using a Gaussian kernel with $\sigma = \frac{1}{5}$ semitones (1 bin) to obtain $\widetilde{Y}$.
Minimization of $\mathcal{L}_{spv}$ represents the classic training objective for supervised MPE and is used as the primary training signal within our framework.

\subsection{Self-Supervised Techniques}\label{sec:self_supervised_techniques}
\subsubsection{Invariance \& Equivariance Objectives}\label{sec:invariance_equivariance_objectives}
We further define two classes of self-supervised objectives, adapted from \cite{cwitkowitz2024toward}, based on pitch-invariant and pitch-equivariant properties.
Under our framework, these objectives are meant to encourage the model to implicitly encode various properties of pitch.
A pitch-invariant transformation $t_{iv}(\cdot)$ performs some manipulation of $X_{\mathcal{H}}$ that ideally should not affect the predicted multi-pitch salience-gram $\hat{Y}$.
These transformations can be used to formulate invariance-based losses of the form
\begin{equation}
\label{eq:invariance_loss}
\mathcal{L}_{iv} = \frac{1}{N} \sum^{N - 1}_{n = 0} \sum^{K - 1}_{k = 0}
\mathcal{B}\hspace{-0.15em}\left( \mathcal{F}\hspace{-0.15em}\left( t_{iv}\hspace{-0.15em}\left( X_{\mathcal{H}} \right)\right)\hspace{-0.25em}[k, n], \hat{Y}[k, n] \right).
\end{equation}
While the relative strength of energy at harmonic frequencies is primarily what influences timbre, pitch is determined by an actual or implied F0.
As such, we simulate pitch-invariant timbral transformations $t_{iv-t}$ by applying random parabolic equalization curves $u[k] = 1 - 2 \alpha (k - \beta)^2$ \cite{abesser2021jazz}, where $\beta \in [0, K - 1]$ and $\alpha \in [0, \frac{1}{(K - 1)^2}]$ are sampled uniformly, to each frame and channel of $X_{\mathcal{H}}$, to define a timbre-invariance loss $\mathcal{L}_{iv-t}$.
Similarly, there is no discernible pitch associated with non-harmonic sounds, which nonetheless make up an important aspect of music (\textit{i.e.}, percussion).
As such, we create musically-relevant pitch-invariant transformations $t_{iv-p}$ by randomly sampling and superimposing percussive audio from the Expanded Groove MIDI Dataset (E-GMD) \cite{callender2020improving} (at the waveform-level) with volume $v \in [0, 1]$, sampled uniformly, to define a percussion-invariance loss $\mathcal{L}_{iv-p}$.

Conversely, a pitch-equivariant transformation $t_{ev}(\cdot)$ performs some manipulation of $X_{\mathcal{H}}$ that ideally should correspond to a parallel manipulation of the predicted multi-pitch salience-gram $\hat{Y}$.
These transformations can be used to formulate equivariance-based losses of the form
\begin{equation}
\label{eq:equivariance_loss}
\mathcal{L}_{ev} = \frac{1}{N} \sum^{N - 1}_{n = 0} \sum^{K - 1}_{k = 0}
\mathcal{B}\hspace{-0.15em}\left( \mathcal{F}\hspace{-0.15em}\left( t_{ev}\hspace{-0.15em}\left( X_{\mathcal{H}} \right)\right)\hspace{-0.25em}[k, n], t_{ev}\hspace{-0.15em}\left( \hat{Y}[k, n] \right)\hspace{-0.15em}\right).
\end{equation}
The HCQT spectrograms fed into our model and the corresponding expected multi-pitch salience-grams have an equivariant relationship for various geometric transformations.
These include vertical translations, which correspond to a pitch shift of $\frac{\Delta_{k}}{5}$ semitones, horizontal translations, which correspond to a time delay of $\frac{4\Delta_{n}}{N}$ seconds, and horizontal stretching, which corresponds to a speed-up by a factor of $\gamma$.
We perform random pitch-equivariant transformations $t_{ev-g}$ with uniformly sampled $\Delta_{k} \in [-b_{oct}, b_{oct}]$ bins, $\Delta_{n} \in [-\frac{N}{4}, \frac{N}{4}]$ frames, and $\gamma \in [0.5, 2]$\footnote{Sampled uniformly from $[0.5, 1]$ and $[1, 2]$ in equal proportion.} to define a geometric-equivariance loss $\mathcal{L}_{ev-g}$.

While all of these invariance and equivariance properties can be learned implicitly to some degree through a supervised objective or fully convolutional inductive bias, an explicit training signal can lead to less overfitting.
Moreover, these techniques can broaden the training data and introduce previously unseen elements such as percussion.

\subsubsection{Energy-Based Stimulus}\label{sec:energy_based_stimulus}
Objectives based on the losses from Sec. \ref{sec:invariance_equivariance_objectives} can be vulnerable to trivial solutions, \textit{e.g.} uniformly inactive or active multi-pitch salience-grams.
One way to protect against such degeneration is to inject some sort of energy-based stimulus.
The ground-truth and its derivative $\widetilde{Y}$ are one form of stimulus that can prevent collapse, but of course they are not always available.
In lieu of ground-truth, a loss leveraging energy-based targets can be formulated as
\begin{equation}
\label{eq:energy_loss}
\mathcal{L}_{eg} = \frac{1}{N} \sum^{N - 1}_{n = 0} \sum^{K - 1}_{k = 0} \mathcal{B}\hspace{-0.15em}\left( \hat{Y}[k, n], \widetilde{X}[k, n] \right),
\end{equation}
where $\widetilde{X}^{(lin)} = \sum^{5}_{h = 1} \frac{1}{h^4} X_{h}^{(lin)}$ represents a weighted average of $X_{\mathcal{H}}$ across harmonic channels, computed in linear-scale and converted to decibel-scale.
Note that (\ref{eq:energy_loss}) is a simplification of the harmonic and support loss used originally in \cite{cwitkowitz2024timbre}.
While this loss will protect against trivial solutions, the target $\widetilde{X}[k, n]$ by nature is quite coarse and contains many false alarms (see \cite{cwitkowitz2024timbre}).
A simple improvement is to induce sparsity through another loss:
\begin{equation}
\label{eq:sparsity_loss}
\mathcal{L}_{spr} = \frac{1}{N} \sum^{N - 1}_{n = 0} \sum^{K - 1}_{k = 0}
\left| \hat{Y}[k, n] \right|.
\end{equation}
In practice, $\mathcal{L}_{eg}$ and $\mathcal{L}_{spr}$, if computed at all, are coupled and computed only for predictions without ground-truth.

\section{Experiments}\label{sec:experiments}
\input{tables/results_main}

In this section, we detail our experimental setup and our initial investigation into the joint training paradigm.

\subsection{Training \& Evaluation Details}\label{sec:training_evaluation_details}
We train and validate the model in each experiment on URMP \cite{li2018creating} following the splits proposed in \cite{gardner2021mt3}.
Training is conducted on batches of $4$ second excerpts using AdamW optimizer \cite{loshchilov2019decoupled} with batch size $8$ and learning rate $0.0005$ for the encoder and $0.00025$ for the decoder.
Only one excerpt per track is sampled over the course of each epoch.
In experiments with self-supervision on additional data, the batch size is expanded to accommodate extra samples without reducing the amount of supervision.
The supervised objective (\ref{eq:supervised_loss}) is computed and averaged across supervised samples, whereas the self-supervised objectives (\ref{eq:invariance_loss}-\ref{eq:sparsity_loss})
are computed and averaged across all samples within each batch. Since (\ref{eq:supervised_loss}-\ref{eq:energy_loss}) are formulated using BCE, they all operate on roughly the same numerical scale.
Learning rate warmup is applied over the first $100$ epochs of training, and gradient clipping with an $L_2$-norm of $1.0$ is applied to improve training stability.
The final model for each experiment is chosen as the checkpoint with the maximum $F_1$-score on the validation set across $2500$ epochs.

We evaluate on several MPE and AMT datasets, including Bach10 \cite{duan2010multiple}, Su \cite{su2016escaping}, TRIOS \cite{fritsch2012master}, the ten-track test set of MusicNet \cite{thickstun2017learning}, and GuitarSet \cite{quingyang2018guitarset}.
We utilize the community-standard \texttt{mir\_eval} package \cite{raffel2014mir_eval} to compute precision ($\mathit{P}$), recall ($\mathit{R}$), and $f_1$-score ($\mathit{F_1}$).
Multi-pitch estimates are generated by performing local peak-picking on the output multi-pitch salience-grams and thresholding at 0.5.
The final results are computed by averaging across all tracks within an individual dataset.

\subsection{Baselines}\label{sec:baseline}
We compare results for our experiments to several supervised CNN-based approaches to MPE.
\textbf{Deep-Salience} \cite{bittner2017deep} feeds an HCQT spectrogram with harmonics $\mathcal{H}$ and 5 bins per semitone into several convolutional layers to produce a multi-pitch salience-gram.
It is functionally similar to our framework under a supervised-only setting.
The model was trained on a private subset of multitrack mixtures (including percussion) from MedleyDB \cite{bittner2014medleydb}.
\textbf{Basic-Pitch} \cite{bittner2022lightweight} is similar to Deep Salience, performing MPE at 3 bins per semitone, but employs a more shallow network and further estimates pitch and onset activations at 1 bin per semitone to generate note predictions.
The input to the model is an approximation of the HCQT.
\footnote{Note that Basic-Pitch utilizes 7 harmonics and a sub-harmonic.}
Data augmentation techniques including additive noise, equalization, and reverb simulation are also utilized.
The model was trained on portions of several medium-to-large-sized datasets, including GuitarSet \cite{quingyang2018guitarset}.
\textbf{PUnet:XL} \cite{weiss2022comparing} is a method drawing inspiration from the idea of pre-stacking AMT models with a U-Net \cite{pedersoli2020improving}.
It processes fixed-length windows of a 6-octave HCQT with harmonics $\mathcal{H}$ and 3 bins per semitone, and makes predictions at 1 bin per semitone.
The model is trained on MusicNet \cite{thickstun2017learning} and incorporates an auxiliary task of degree-of-polyphony estimation at the latent layer and data augmentation techniques including transposition (similar to our geometric-equivariance objective), tuning manipulation, additive noise, and equalization following \cite{abesser2021jazz}.
\textbf{Timbre-Trap} \cite{cwitkowitz2024timbre} is a 2D autoencoder designed to perform MPE and audio synthesis jointly based on a simple conditioning mechanism at the latent space.
The backbone architecture is nearly identical to that of our framework, barring the modifications noted in Sec. \ref{sec:model_features}.
However, in the original framework the model received both the real and imaginary part (as separate channels) of an invertible complex CQT \cite{holighaus2012framework} as input.
Timbre-Trap was trained on URMP \cite{li2018creating} following the same splits used in this work.
We follow the same post-processing steps described in Sec. \ref{sec:training_evaluation_details} to evaluate the baseline models,\footnote{Adopting the original hyperparameters, we threshold Deep-Salience and Basic-Pitch at 0.3, and PUnet:XL (without peak-picking) at 0.4.}
and provide the results at the top of Table \ref{tab:results_main}.

\subsection{Joint Training Paradigm}\label{sec:joint_training_paradigm}
We first conduct an initial set of experiments evaluating the invariance- and equivariance-based self-supervised objectives under closed training conditions on URMP \cite{li2018creating}.
In particular, we experiment with the supervised objective in isolation, the supervised objective with each invariance- and equivariance-based objective independently, and all of these objectives together: $\mathcal{L}_{total} = \mathcal{L}_{spv} + \mathcal{L}_{iv-t} + \mathcal{L}_{iv-p} + \mathcal{L}_{ev-g}$.
The results are given at the bottom of Table \ref{tab:results_main}.
The first thing to note is that overall our framework under the supervised-only setting achieves results comparable to or better than each of the baselines for several datasets.
This is especially true with respect to Timbre-Trap \cite{cwitkowitz2024timbre}, which is arguably the most comparable due to its similar architecture and identical training data.
However, PUnet:XL \cite{weiss2022comparing} appears to offer a significant advantage for AMT datasets (\textit{i.e.}, Su, TRIOS, and MusicNet) since it was trained on such data to predict pitch directly at the note-level.
Next, we can observe that each self-supervised objective has the potential to improve performance on one or more datasets.
The timbre-invariance objective is least effective and has a mixed effect across datasets, which is somewhat contrary to what has been observed under the fully self-supervised context \cite{cwitkowitz2024toward}.
It is possible that the property of timbre-invariance is already strongly indicated by the supervised objective.
The percussion-invariance objective is moderately beneficial, even for some datasets without percussion.
Out of all the evaluation datasets, there is only one track in TRIOS \cite{fritsch2012master} that has percussion.
However, even non-percussive audio can have percussive elements, \textit{i.e.} originating from playing notes on certain instruments.
The geometric-equivariance objective is most effective and yields a significant improvement across all datasets.
This is likely due to the model capturing harmonic relationships explicitly and efficiently by leveraging the shift-invariance exhibited by the HCQT representation.
Combining the supervised objective with all invariance- and equivariance-based objectives produces the best performance, suggesting that each contributes distinctly to robustness.
Given that our framework was trained with such a small of amount of audio (\textit{i.e.}, 1-2 hours), these results are quite remarkable, especially when considering performance on data unseen during training (\textit{i.e.}, GuitarSet).

\subsection{Self-Supervision on Additional Data}\label{sec:self_supervision_on_additional_data}
\begin{figure*}[t]
\centering
\includegraphics[width=0.95\linewidth]{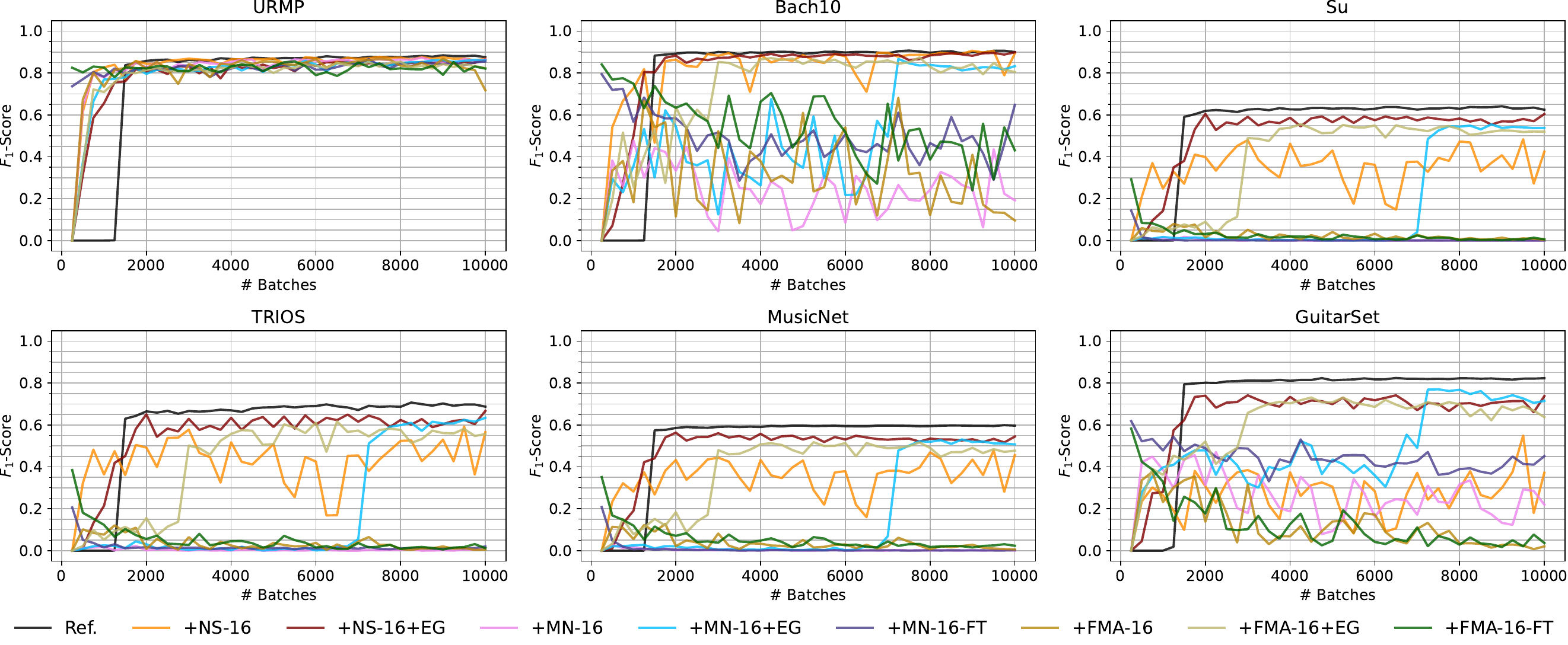}
\caption{Performance over the course of training for experiments leveraging an additional 16 samples per batch from NSynth \cite{engel2017neural}, MusicNet \cite{thickstun2017learning}, and FMA \cite{defferrard2016fma} for self-supervision. EG - Energy-based objectives. FT - Fine-tuning scheme.}
\label{fig:additional}
\end{figure*}

\begin{figure*}[t]
\centering
\includegraphics[width=0.99\linewidth]{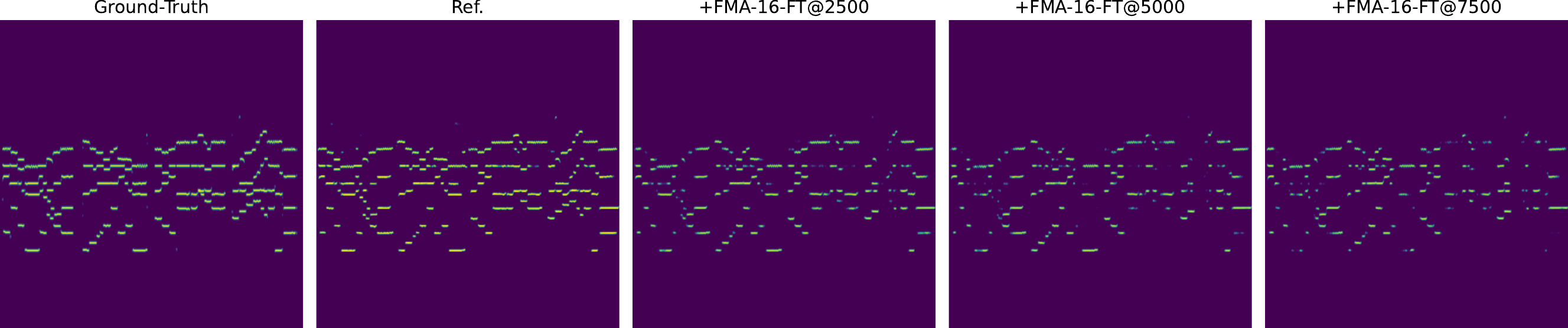}
\caption{Ground-truth, baseline, and intermediate predictions at 25\%, 50\%, and 75\% of the duration of the experiment with additional self-supervision on FMA \cite{defferrard2016fma} under fine-tuning scheme for track \texttt{01-AchGottundHerr} of Bach10 \cite{duan2010multiple}.}
\label{fig:degeneration}
\end{figure*}

Given the success with integrating self-supervised objectives into our supervised framework, it is reasonable to question whether the same objectives could be applied to more general music datasets lacking multi-pitch annotations.
Indeed, training a model to maintain pitch-invariant and pitch-equivariant properties over a broader collection of data could be one potential way to circumvent issues with low data availability for MPE.
In this vein, we conduct additional experiments under the joint training paradigm where additional data is included in each batch for self-supervision only.
Specifically, we repeat the experiment combining all objectives (denoted \texttt{Ref.}), but with an additional 16 samples per batch which only influence the invariance- and equivariance-based losses.
The datasets we use for additional samples represent different music domains, \textit{i.e.}, simple synthetic monophonic data (NSynth \cite{engel2017neural}), recordings of classical music mixtures (MusicNet \cite{thickstun2017learning}), and high-quality production-level audio (FMA \cite{defferrard2016fma}).
We extract samples from the training splits of NSynth and MusicNet, and the large (30-second clip) variant of FMA.

Surprisingly, these experiments all exhibit undesirable behavior: performance for the URMP \cite{li2018creating} remains stable and consistent with experiments from Sec. \ref{sec:joint_training_paradigm}, but performance for the other datasets collapses.
The performance at validation checkpoints over the course of training is presented in Fig. \ref{fig:additional}.
Upon closer inspection, we found that the model predictions degenerate to a trivial solution (blank predictions) for all datasets except for URMP.
As such, we repeated each experiment with the energy-based objectives (+EG) from Sec. \ref{sec:energy_based_stimulus} on the data used for self-supervision.
Although this does prevent collapse, it ultimately still leads to degraded performance.
Finally, we experiment with initializing the model with the weights from the best validation checkpoint of \texttt{Ref.} and fine-tuning with $\frac{1}{5}$ the learning rate (-FT).
However, the two-stage fine-tuning paradigm still exhibits the same behavior.

\section{Discussion}\label{sec:discussion}

In this section, we investigate the phenomenon uncovered in Sec. \ref{sec:self_supervision_on_additional_data} further and conduct several follow-up experiments in an effort to identify the underlying problem.

\subsection{Overfitting \& Degeneration}\label{sec:overfitting_degeneration}
\begin{figure*}[t]
\centering
\includegraphics[width=0.95\linewidth]{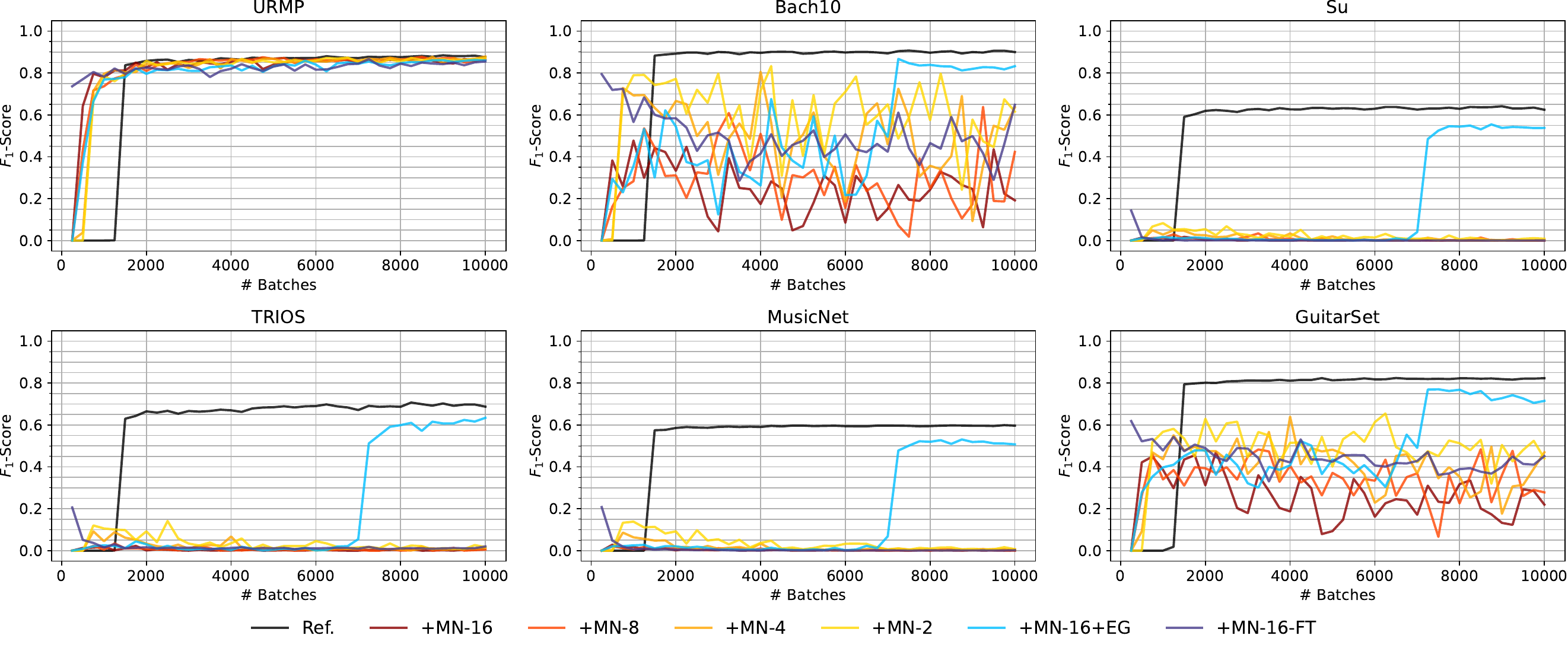}
\caption{Performance over the course of training for experiments varying amount of self-supervision on MusicNet \cite{thickstun2017learning}.}
\label{fig:musicnet}
\end{figure*}

\begin{figure*}[t]
\centering
\includegraphics[width=0.95\linewidth]{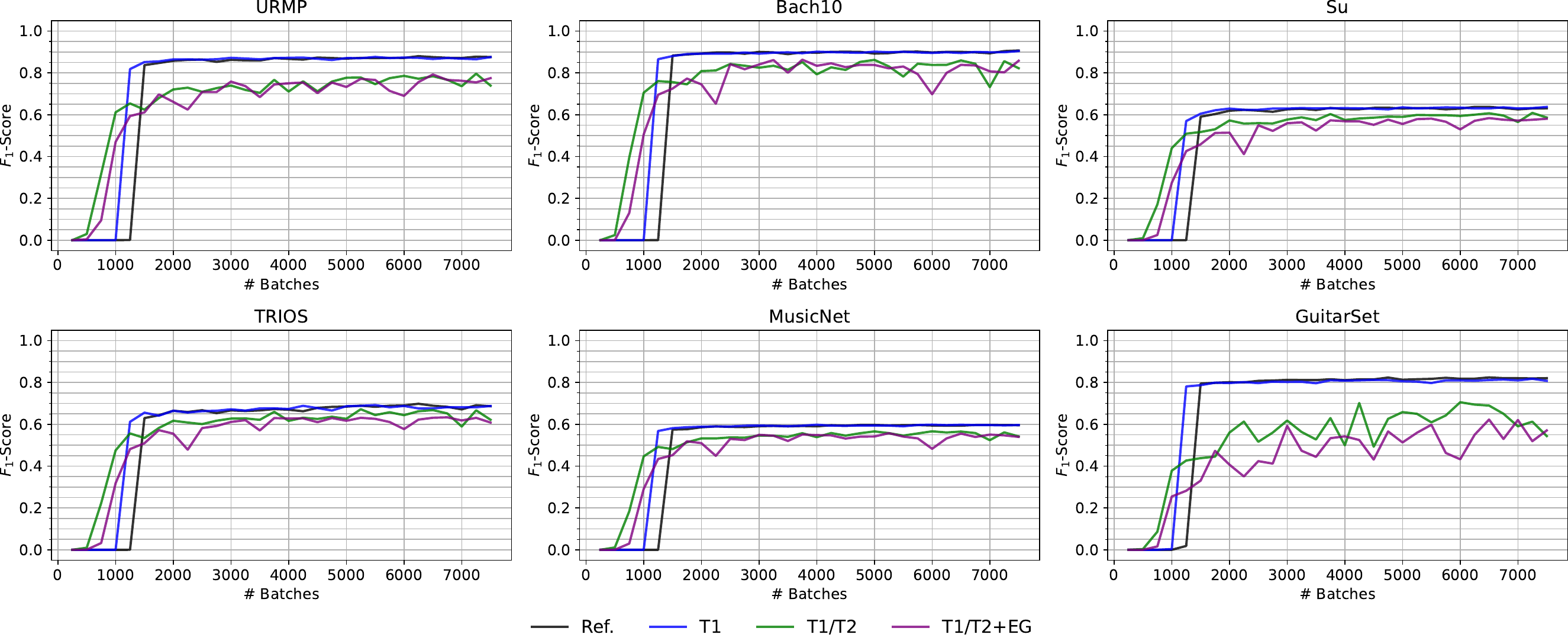}
\caption{Performance over the course of training for experiments which use a reduced portion (T1) of our URMP \cite{li2018creating} training set for supervised training with additional self-supervision on the remaining 10 non-overlapping samples (T2).}
\label{fig:urmp_t1/t2}
\end{figure*}

In order to illustrate and characterize the problem of degeneration, Fig. \ref{fig:degeneration} shows predictions for a single sample from \texttt{Ref.} along with predictions for the same sample at 25\%, 50\%, and 75\% of the full duration of fine-tuning on FMA.
It can clearly be seen that the strength of predictions (\textit{i.e.}, recall) decreases over time, indicating that there is a trend towards a trivial solution.
This also suggests that for the regular experiments without fine-tuning, the model is always struggling to move past the trivial solution.

Next, to examine whether the role of self-supervised learning was too extreme, we repeat the MusicNet \cite{thickstun2017learning} experiments with an exponentially decreasing amount of samples for self-supervision only (\textit{i.e.}, with 8, 4, and 2 additional samples).
The performance of these experiments at validation checkpoints is plotted in Fig. \ref{fig:musicnet}.
Interestingly, there is a noticeable relationship between the amount of samples for self-supervision only and the severity of degeneration.
Moreover, the degeneration on MusicNet \cite{thickstun2017learning} itself is quite prominent and also exhibits this relationship.

\subsection{Training Distributions}\label{sec:training_distributions}
In order to see whether there is an issue regarding mismatch between the distribution of the data used for supervision and that of the data used for self-supervision only, we further extract 10 samples from the URMP \cite{li2018creating} training set spanning multiple degrees of polyphony.
We denote this split as URMP-T2 and the remainder as URMP-T1.
We then re-run the reference experiment using only URMP-T1 for training, and then again with additional self-supervised learning and no corresponding supervision on URMP-T2.
For completeness, we further run this experiment applying the energy-based objectives to URMP-T2.
The performance for these experiments is plotted in Fig. \ref{fig:urmp_t1/t2}.

Relative to \texttt{Ref.}, performance barely decreases when only URMP-T1 is used for training, which is quite interesting by itself.
More importantly, when conducting additional self-supervision on the 10 samples of URMP-T2, which follow a very similar distribution to URMP-T1, there is still a moderate decrease in performance.
The only difference between these two experiments is that each of the self-supervised losses are averaged over the samples from both URMP-T1 and URMP-T2 for each batch instead of only URMP-T1.
It is worth noting that the collapse here happens to be less extreme than when training with self-supervision on other datasets with more samples (see Fig. \ref{fig:musicnet}).
Furthermore, the degradation on URMP and Bach10 is actually more extreme than what we observed when training on NSynth (+NS-16), but we note that this could also be explained by less supervision on URMP.
Another interesting observation is that the energy-based objectives actually further degrade performance, likely due to conflicting with the supervised objective on the same distribution.

\subsection{Mitigating Degeneration}\label{sec:mitigating_degeneration}
Given all of our observations, it appears the underlying issue is too strong of a pull towards the trivial solution for the non-supervised data, irrespective of its distribution.
As it stands, self-supervision without corresponding supervision essentially pushes the model to degenerate on data following a certain distribution.
If the distribution of data used for self-supervision is close to that of the supervised data, it will by extension hurt performance on the supervised data.
If the distribution of data used for self-supervision is distinct from that of the supervised data, it will have less of an effect on the performance for data outside the distribution.
It is unclear whether these interactions would persist if larger and more diverse data were used for supervision.

Degeneration also seems to be unique to MPE, since self-supervised methods for monophonic pitch estimation \cite{gfeller2020spice, riou2023pesto} rely on the inductive bias of monophony and formulate their objectives using categorical cross-entropy.
A solution for the polyphonic setting may require some sort of objective that enforces the existence of content in the predictions, to protect against the trivial solution.
The energy-based objectives are one such protection, but they evidently remove too much flexibility and lead to worse predictions.
Despite these current challenges, we still believe that self-supervised learning holds promise for advancing MPE.

\section{Conclusion}
We have demonstrated that self-supervised objectives can substantially improve upon the standard supervised training paradigm for MPE.
However, in attempting to extend self-supervised learning beyond the distribution of data that is already grounded with supervised learning, we encounter issues whereby our model simultaneously overfits to the distribution of the supervised training data while degenerating on the distribution of the self-supervised training data.
Self-supervised objectives utilizing energy-based targets can protect against degeneration, but these are too inflexible.
Fine-tuning cannot circumvent the problem either.
Moreover, we show that degeneration persists even when the supervised and self-supervised training data are taken from the same distribution.
We conclude with several remarks and ideas toward overcoming highlighted issues.

\onecolumn
\begin{multicols}{2}

\section{Acknowledgments}
This work is supported by National Science Foundation (NSF) grant No. 2222129 and synergistic activities funded by NSF grant DGE-1922591.

\bibliography{ssmpe}
\end{multicols}

\end{document}

%% file: tables/results_main.tex
\begin{table*}[!t]
\caption{Comparison of precision ($\mathit{P}$), recall ($\mathit{R}$), and $f_1$-score ($\mathit{F_1}$) (in percentage points) over several MPE and AMT datasets for baseline methods\textsuperscript{*} and experiments conducted using the proposed framework with self-supervised objectives.}
\label{tab:results_main}
\footnotesize
\setlength{\tabcolsep}{0.6em}
\renewcommand{\arraystretch}{1.35}
\centering
\begin{threeparttable}
\begin{tabular}{||c||c|c|c||c|c|c||c|c|c||c|c|c||c|c|c||}
    \hline
    & \multicolumn{3}{c||}{\textbf{Bach10}} & \multicolumn{3}{c||}{\textbf{Su}} & \multicolumn{3}{c||}{\textbf{TRIOS}} & \multicolumn{3}{c||}{\textbf{MusicNet}} & \multicolumn{3}{c||}{\textbf{GuitarSet}} \\
    \hline
    \textbf{Method} & $\mathit{P}$ & $\mathit{R}$ & $\mathit{F_1}$ & $\mathit{P}$ & $\mathit{R}$ & $\mathit{F_1}$ & $\mathit{P}$ & $\mathit{R}$ & $\mathit{F_1}$ & $\mathit{P}$ & $\mathit{R}$ & $\mathit{F_1}$ & $\mathit{P}$ & $\mathit{R}$ & $\mathit{F_1}$ \\
    \hline
    \hline
    Deep-Salience \cite{bittner2017deep} & $86.0$ & $61.0$ & $71.3$ & $74.1$ & $47.9$ & $57.1$ & $\mathbf{92.3}$ & $39.4$ & $54.2$ & $63.0$ & $48.1$ & $53.3$ & $77.7$ & $70.6$ & $72.2$ \\
    \hline
    Basic-Pitch \cite{bittner2022lightweight} & $90.2$ & $75.7$ & $82.2$ & $54.2$ & $44.1$ & $47.5$ & $88.2$ & $44.2$ & $57.9$ & $50.6$ & $42.1$ & $45.7$ & \textcolor{gray}{$\mathbf{80.9}$} & \textcolor{gray}{$75.9$} & \textcolor{gray}{$77.7$} \\
    \hline
    PUnet:XL \cite{weiss2022comparing} & $88.2$ & $77.6$ & $82.5$ & $\mathbf{76.2}$ & $\mathbf{69.8}$ & $\mathbf{71.8}$ & $89.6$ & $51.3$ & $64.8$ & $\mathbf{77.6}$ & $\mathbf{67.6}$ & $\mathbf{72.0}$ & $74.7$ & $55.2$ & $62.4$ \\
    \hline
    Timbre-Trap \cite{cwitkowitz2024timbre} & $81.2$ & $84.2$ & $82.6$ & $52.1$ & $53.0$ & $51.4$ & $69.4$ & $49.7$ & $56.8$ & $44.1$ & $57.3$ & $48.7$ & $48.6$ & $75.6$ & $58.0$ \\
    \hline
    \hline
    $\mathcal{L}_{spv}$ & $88.8$ & $83.1$ & $85.8$ & $61.0$ & $47.7$ & $52.1$ & $82.7$ & $48.0$ & $59.4$ & $53.8$ & $54.7$ & $53.6$ & $70.2$ & $72.9$ & $69.8$ \\
    \hline
    $\mathcal{L}_{spv} + \mathcal{L}_{iv-t}$ & $88.5$ & $81.7$ & $84.9$ & $62.5$ & $47.1$ & $52.3$ & $84.4$ & $46.4$ & $58.8$ & $54.9$ & $54.3$ & $53.9$ & $75.4$ & $70.1$ & $70.6$ \\
    \hline
    $\mathcal{L}_{spv} + \mathcal{L}_{iv-p}$ & $88.0$ & $84.1$ & $85.9$ & $57.3$ & $52.5$ & $53.4$ & $81.0$ & $50.1$ & $60.6$ & $50.4$ & $58.9$ & $53.8$ & $70.9$ & $78.0$ & $73.2$ \\
    \hline
    $\mathcal{L}_{spv} + \mathcal{L}_{ev-g}$ & $91.5$ & $\mathbf{88.6}$ & $\mathbf{90.0}$ & $64.9$ & $62.9$ & $62.9$ & $90.7$ & $57.9$ & $69.8$ & $58.0$ & $62.9$ & $59.8$ & $79.7$ & $80.4$ & $79.3$ \\
    \hline
    $\mathcal{L}_{total}$ (\texttt{Ref.}) & $\mathbf{92.1}$ & $88.0$ & $90.0$ & $65.0$ & $65.0$ & $64.1$ & $91.0$ & $\mathbf{58.4}$ & $\mathbf{70.2}$ & $55.8$ & $65.1$ & $59.6$ & $80.5$ & $\mathbf{82.3}$ & $\mathbf{80.9}$ \\
    \hline
\end{tabular}
\begin{tablenotes}\footnotesize
\item[*] Grayed values indicate a portion of the test data was used for training.
\end{tablenotes}
\end{threeparttable}
\end{table*}